\documentclass[runningheads,orivec]{llncs}
\usepackage[T1]{fontenc}
\usepackage{newtxtext}
\usepackage{amsmath}
\usepackage[varvw]{newtxmath} \usepackage{booktabs,array}
\usepackage{graphicx} \usepackage{algorithm,algpseudocode}
\usepackage{tikz}
\usetikzlibrary{arrows.meta,positioning,fit,calc,shapes.geometric}
\algrenewcommand{\algorithmicrequire}{\textbf{Input:}}
\algrenewcommand{\algorithmicensure}{\textbf{Output:}}
\usepackage{microtype}
\usepackage{needspace}
\makeatletter
\let\llncsfnsymbol\@fnsymbol
\let\llncsmakefntext\@makefntext
\let\llncsmakefnmark\@makefnmark
\makeatother
\usepackage[bottom]{footmisc}
\makeatletter
\let\@fnsymbol\llncsfnsymbol
\let\@makefntext\llncsmakefntext
\let\@makefnmark\llncsmakefnmark
\makeatother
\usepackage[hidelinks]{hyperref}
\newcommand{\arctn}{\textsc{ArcTN}}
\newcommand{\Mmax}{M_{\max}}
\newcommand{\Fall}{F_{\mathrm{all}}}
\newcommand{\Nslice}{N_{\mathrm{slice}}}

\title{ArcTN: Efficient Tensor-Network Contraction
Planning for Quantum Circuit Simulation}
\author{Zhengyi Liu\inst{1}\thanks{Equal contribution.} \and
Li Liu\inst{2}\inst{\thefootnote} \and
Songtao Liu\inst{1} \and
Xianglong Liu\inst{1} \and
Xiangzhen Zhou\inst{1,4}\thanks{Corresponding authors: \email{zhouxiangz@qq.com}, \email{chenbl3@hygon.cn}, \email{guanj@ios.ac.cn}.} \and
Mingyu Huang\inst{1} \and
Dingchao Gao\inst{1} \and
Shenggang Ying\inst{3} \and
Bolong Chen\inst{2}\inst{\thefootnote} \and
Ji Guan\inst{3}\inst{\thefootnote}
}
\authorrunning{Z. Liu et al.}
\titlerunning{ArcTN: Tensor-Network Contraction Planning} \authorrunning{}
\institute{Arclight Quantum Computing Inc. \and
HYGON \and
Institute of Software, Chinese Academy of Sciences \and
Nanjing Tech University}
\authorrunning{Z. Liu et al.}
\begin{document}
\maketitle

\begin{abstract}
Efficient tensor-network simulation of quantum circuits critically
depends on finding contraction orders that minimize computational
cost without excessive optimization overhead or large intermediate
tensors. We present \arctn{}, a contraction-order optimization
framework that addresses these challenges through adaptive search
and memory-aware slicing. \arctn{} provides two complementary search presets, \emph{Light} and \emph{Heavy},
that combine diversified initial orders, adaptive randomized-greedy
search, selective subtree reconfiguration, and parallel tempering
with stage-specific stopping criteria.
\emph{Light} prioritizes low planning overhead, whereas \emph{Heavy} performs
more extensive refinement to obtain lower-cost contraction orders.
The framework also supports \emph{Fixed} and \emph{Dynamic} slicing to control
intermediate tensor sizes, together with numerical execution
using optimized contraction orders. We evaluate \arctn{} on QC32, a benchmark comprising 32
quantum-circuit tensor networks, against established contraction
optimizers.
On the 30 circuits completed by all eight main configurations,
\arctn{} \emph{Light} reduces contraction FLOPs by 61.32\% with a
$1.62\times$ planning-time speedup over
{Cotengra Auto/CMA-ES}.
\arctn{} \emph{Heavy} reduces FLOPs by 43.83\% with a
$7.21\times$ planning-time speedup over
{Cotengra AutoHQ/Optuna}, and by 11.85\% with a
$1.33\times$ speedup over {OMECO TreeSA-8}.
For memory-constrained contractions, \emph{Dynamic} slicing satisfies
the prescribed intermediate-tensor size limits in all 90 runs on the 18 common scalar-output instances and reduces total sliced FLOPs by 31.99\%
relative to \emph{Fixed} slicing, at the cost of additional planning time.
Experiments with a common NumPy executor further demonstrate
that \arctn{} \emph{Heavy} achieves the lowest combined planning
and execution time among five evaluated configurations on
their shared successful cases.
On 672 additional networks built from 320 MQT Bench circuits for amplitude, expectation-value, and state tasks, \arctn{} \emph{Heavy} obtains the lowest geometric-mean FLOPs on the 142 large networks while planning $3.1\times$ and $31\times$ faster than Cotengra AutoHQ and Hyper128.

\end{abstract}

\section{Introduction}
\label{sec:introduction}

Classical simulation supports the development of quantum algorithms,
the validation of quantum devices, and the study of quantum
dynamics~\cite{bayraktar2023cuquantum}. Many simulation tasks require
only selected quantities, such as a circuit amplitude or an expectation
value, rather than the complete quantum state. Tensor networks (TNs)
provide a common representation for these computations. Tensors encode
operations and states, and contraction evaluates the requested quantity
by multiplying tensor entries and summing over non-output
indices~\cite{markov2008,gray2021}. This representation allows a
simulator to exploit the structure of the circuit and the requested
output without explicitly constructing the full state vector.

However, a tensor-network representation does not by itself ensure
efficient simulation. The same network can be contracted in many
different orders, with substantially different arithmetic costs and
intermediate tensor sizes~\cite{markov2008,pfeifer2014,gray2021}.
An unsuitable order can make a contraction unnecessarily expensive
or produce tensors too large to store. Efficient simulation therefore
depends on \emph{contraction planning}, which selects an order that reduces
computational cost while accounting for the resources needed to find
and execute it.

The difficulty is that contraction planning must balance three coupled
concerns. First, the returned order should have low computational cost.
Second, finding that order should not require excessive planning time.
More extensive search may improve an order, but its additional cost
must be justified by the execution savings, either for one expensive
contraction or across repeated contractions with the same network
structure. Third, the plan must accommodate tensor-size constraints.
Slicing selected indices splits a contraction into smaller tasks and
reduces individual tensor sizes, but can increase the total computation
across those tasks~\cite{gray2021}. These concerns cannot be treated
independently, because the order determines which tensors become large, while
slicing changes the costs of alternative orders. A good unsliced order
is therefore not necessarily a good starting point for a low-cost
sliced computation.

Existing optimizers, including Cotengra~\cite{gray2021}, OMECO~\cite{omecorust}, and TNCO~\cite{tnco}, provide
techniques for constructing and improving contraction
orders. The practical challenge is not
simply to search more extensively, but to coordinate these techniques.
Which initial orders should be explored, which candidates deserve
further refinement, when should a search stage stop, and how should
the order be revised when slicing is required? Moreover, improvements
in modeled operation count must be assessed alongside actual
execution time, since fewer operations do not by themselves guarantee
faster execution on a particular backend.

We present \arctn{}, a contraction-planning tool that addresses these
concerns through adaptive search and tensor-size-aware slicing.
Its two complementary search presets, \emph{Light} and \emph{Heavy},
organize established construction and refinement techniques with
different levels of planning effort. Both use the cross-entropy method
to adapt random-greedy candidate generation, retain multiple starting
orders, and selectively refine candidates through subtree
reconfiguration and parallel tempering. Retaining multiple starting
points allows refinement to explore orders whose initial costs may
not reflect their eventual quality. Stage-specific stopping criteria,
together with overall time- and rate-based checks, control the effort
spent on construction and refinement. \emph{Light} prioritizes low
planning overhead, whereas \emph{Heavy} devotes more effort to
obtaining lower-cost orders.

For tensor-size constraints, \arctn{} provides \emph{Fixed} and
\emph{Dynamic} slicing. \emph{Fixed} selects sliced indices while
preserving the chosen contraction order. \emph{Dynamic} additionally
alternates slice selection with local order refinement, retaining
the feasible plan with the lowest modeled total cost across all
slices. This allows the order to be reconsidered as slicing changes
the computation, rather than treating slicing solely as a
postprocessing step. Both modes target a user-specified bound on
the size of each contraction result within a slice.

\arctn{} implements this workflow in Rust with Python interfaces.
Given a tensor-network description, it validates the input, searches
for an order, optionally constructs a sliced plan, and supports
numerical execution through a native CPU executor or an external
array backend. Execution plans can be retained and reused with new
tensor values without repeating the order search, connecting planning
to both one-off and repeated computations.

\Needspace{6\baselineskip}
In summary, the main contributions of this tool paper are:
\begin{enumerate}
  \item \textbf{A coordinated planning framework.}
  We organize adaptive candidate generation, selective refinement,
  and stopping control into the \emph{Light} and \emph{Heavy} search
  presets, complemented by \emph{Fixed} and \emph{Dynamic} slicing
  for tensor-size constraints.

  \item \textbf{An executable contraction tool.}
  We provide a Rust implementation with Python interfaces that
  connects order search and slicing to numerical execution,
  reusable execution plans, and integration with external
  tensor-network workflows.

  \item \textbf{An evaluation of planning and execution benefits.}
  We evaluate \arctn{} on QC32, the 32 quantum-computing instances
  from Einsum Benchmark~\cite{einsumbenchmark}, and on 672 networks
  built from 320 MQT Bench circuits~\cite{mqtbench}.
  On QC32,
  \emph{Light} reduces modeled FLOPs by 61.32\% with a
  $1.62\times$ planning-time speedup over Cotengra Auto/CMA-ES,
  while \emph{Heavy} reduces FLOPs by 43.83\% with a
  $7.21\times$ speedup over Cotengra AutoHQ/Optuna.
  With a common NumPy executor, \emph{Heavy} also achieves the lowest
  combined planning--execution time among five configurations.
   \emph{Dynamic} reduces total
  sliced FLOPs by 31.99\% relative to \emph{Fixed} for scalar outputs.
  On the large MQT Bench networks, \emph{Heavy} obtains the lowest
  geometric-mean FLOPs while planning $3.1\times$ and $31\times$
  faster than Cotengra AutoHQ and Hyper128.
\end{enumerate}

The remainder of this paper is organized as follows. Section~\ref{sec:problem} introduces tensor networks and formalizes contraction planning, its cost model, and tensor-size constraints. Section~\ref{sec:architecture} presents the architecture and capabilities of \arctn{}, including execution and plan reuse. Section~\ref{sec:methods} describes the contraction-order search framework and the \emph{Fixed} and \emph{Dynamic} slicing methods. Section~\ref{sec:implementation} describes the exact integer cost arithmetic of \arctn{}'s dynamic programs. Section~\ref{sec:evaluation} evaluates planning efficiency, contraction quality, numerical execution, and slicing under tensor-size constraints, and planning on circuits from MQT Bench. Section~\ref{sec:related} discusses related tools, and Section~\ref{sec:conclusion} concludes the paper.

\section{Tensor Networks and the Contraction Planning Problem}
\label{sec:problem}

This section defines tensor-network contraction and the cost of a contraction order, and then states the planning problem without and with a limit on tensor sizes.

\subsection{Tensor Networks and Contraction}
\label{sec:tn}

A tensor-network contraction problem specifies input tensors $\mathcal T=\{T_1,\ldots,T_n\}$, an ordered tuple of index labels $\boldsymbol{\ell}_v$ for each tensor $T_v$, a positive integer dimension $d_e$ for every label $e$, and an ordered tuple $\boldsymbol{o}$ of distinct output labels~\cite{bridgeman2017}. The tuple $\boldsymbol{\ell}_v$ assigns one index label to each axis of the input tensor $T_v$, in axis order. Each label denotes a variable $x_e\in\{0,\ldots,d_e-1\}$, and all axes with that label have dimension $d_e$ and use the same variable. Let $I_v$ be the set of distinct labels in $\boldsymbol{\ell}_v$, let $I=\bigcup_v I_v$, and let $O\subseteq I$ be the set of labels in $\boldsymbol{o}$. The contraction computes
\begin{equation}
  Y_{\mathbf x_{\boldsymbol{o}}}
  =\sum_{\mathbf x_{I\setminus O}}
    \prod_{v=1}^{n}T_v[\mathbf x_{\boldsymbol{\ell}_v}].
  \label{eq:tn}
\end{equation}
Here, $\mathbf x_{\boldsymbol{\ell}_v}$ lists variable values in the axis order of $T_v$. The sum ranges over all assignments to non-output labels, while output labels are retained in the order specified by $\boldsymbol{o}$, yielding a scalar when $O$ is empty. Because $O$ explicitly determines which labels remain in the result, even a label shared by several inputs can be retained. More generally, a label may occur in any number of tensors, and if it appears on multiple axes of the same tensor, those axes use the same variable value, selecting the corresponding diagonal entries.

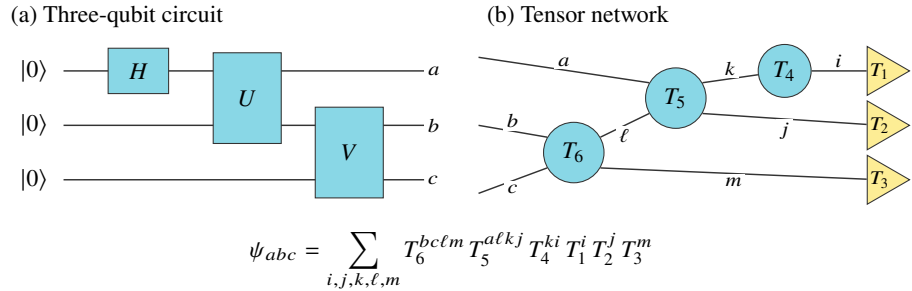
\begin{figure}[htbp]
\definecolor{DiagramCyan}{HTML}{49C2D9}
\definecolor{DiagramYellow}{HTML}{FDED95}
\centering
\begin{tikzpicture}[
  x=0.9mm,y=0.9mm,
  wire/.style={draw=black!80,line width=0.55pt},
  gate/.style={draw=black!80,fill=DiagramCyan!65,minimum width=8mm,minimum height=6mm,inner sep=0.5mm,font=\footnotesize},
  tensor/.style={circle,draw=black!80,fill=DiagramCyan!65,minimum size=7mm,inner sep=0pt,font=\footnotesize},
  state/.style={isosceles triangle,isosceles triangle apex angle=60,draw=black!80,fill=DiagramYellow,minimum size=5.6mm,inner sep=0pt,font=\scriptsize},
  idx/.style={font=\scriptsize,fill=white,inner sep=0.4mm}
]
\node[anchor=west,font=\footnotesize] at (0,30) {(a) Three-qubit circuit};
\draw[wire] (9,22)--(62,22);
\draw[wire] (9,14)--(62,14);
\draw[wire] (9,6)--(62,6);
\node[anchor=east,font=\footnotesize] at (8,22) {$|0\rangle$};
\node[anchor=east,font=\footnotesize] at (8,14) {$|0\rangle$};
\node[anchor=east,font=\footnotesize] at (8,6) {$|0\rangle$};
\node[gate] at (20,22) {$H$};
\node[gate,minimum width=9mm,minimum height=12mm] at (36,18) {$U$};
\node[gate,minimum width=9mm,minimum height=12mm] at (51,10) {$V$};
\node[idx,anchor=west] at (62,22) {$a$};
\node[idx,anchor=west] at (62,14) {$b$};
\node[idx,anchor=west] at (62,6) {$c$};
\begin{scope}[xshift=63mm,yshift=32.4mm]
\node[anchor=west,font=\footnotesize] at (0,-6) {(b) Tensor network};
\node[state] (t1) at (59,-14) {$T_1$};
\node[state] (t2) at (59,-22) {$T_2$};
\node[state] (t3) at (59,-30) {$T_3$};
\node[tensor] (t4) at (45,-14) {$T_4$};
\node[tensor,minimum size=8mm] (t5) at (29,-18) {$T_5$};
\node[tensor,minimum size=8mm] (t6) at (14,-26) {$T_6$};
\draw[wire] (t1)--node[idx,above]{$i$}(t4);
\draw[wire] (t4)--node[idx,above]{$k$}(t5.30);
\draw[wire] (t2)--node[idx,below]{$j$}(t5.330);
\draw[wire] (t5.150)--node[idx,above]{$a$}(0,-12);
\draw[wire] (t5.210)--node[idx,below]{$\ell$}(t6.30);
\draw[wire] (t3)--node[idx,below]{$m$}(t6.330);
\draw[wire] (t6.150)--node[idx,above]{$b$}(0,-22);
\draw[wire] (t6.210)--node[idx,below]{$c$}(0,-32);
\end{scope}
\node[font=\footnotesize] at (66,-6) {$\displaystyle\psi_{abc}=\sum_{i,j,k,\ell,m} T_6^{bc\ell m}\,T_5^{a\ell k j}\,T_4^{k i}\,T_1^{i}\,T_2^{j}\,T_3^{m}$};
\end{tikzpicture}
\caption[A quantum circuit and its tensor-network representation.]{A quantum circuit and its tensor-network representation.}
\label{fig:intro-circuit}
\end{figure}

\begin{figure}[htbp]
\definecolor{DiagramCyan}{HTML}{49C2D9}
\definecolor{DiagramOrange}{HTML}{F47254}
\centering
\begin{tikzpicture}[
  x=1mm,y=1mm,font=\scriptsize,
  wire/.style={draw=black!75,line width=0.5pt},
  tensor/.style={circle,draw=black!80,fill=DiagramCyan!65,minimum size=6mm,inner sep=0pt},
  projection/.style={draw=black!80,fill=DiagramOrange!40,minimum width=7mm,minimum height=4.5mm,inner sep=0.3mm},
  bra/.style={isosceles triangle,isosceles triangle apex angle=60,shape border rotate=180,draw=black!80,fill=DiagramOrange!40,minimum size=4.5mm,inner sep=0pt},
  label/.style={font=\tiny,fill=white,inner sep=0.2mm},
  heading/.style={font=\scriptsize\bfseries,anchor=west}
]
\begin{scope}
\node[heading] at (0,5) {(a) Full state vector};
\node[tensor] (psi) at (6,-8) {$\psi$};
\draw[wire] (psi.35)--node[label,above]{$i$}(24,-2);
\draw[wire] (psi.325)--node[label,below]{$j$}(24,-14);
\node at (14,-21) {$\psi_{ij}$};
\end{scope}
\begin{scope}[xshift=30mm]
\node[heading] at (0,5) {(b) Single amplitude};
\node[tensor] (psi) at (23,-8) {$\psi$};
\node[bra] (pi) at (10,-2) {};
\node[bra] (pj) at (10,-14) {};
\node[anchor=east,inner sep=0.5mm] at (pi.west) {$\langle x_0|$};
\node[anchor=east,inner sep=0.5mm] at (pj.west) {$\langle x_1|$};
\draw[wire] (psi.145)--node[label,above]{$i$}(pi.lower side);
\draw[wire] (psi.215)--node[label,below]{$j$}(pj.lower side);
\node at (14,-21) {$\langle x_0x_1|\psi\rangle$};
\end{scope}
\begin{scope}[xshift=60mm]
\node[heading] at (0,5) {(c) Expectation value};
\node[tensor] (psi) at (4,-8) {$\psi$};
\node[tensor] (bra) at (25,-8) {$\psi^*$};
\node[projection,minimum width=5mm] (op) at (14.5,-2) {$O$};
\draw[wire] (psi.35)--node[label,above]{$i$}(op.west);
\draw[wire] (op.east)--node[label,above]{$i'$}(bra.145);
\draw[wire] (psi.325)--(8,-14)--node[label,below]{$j$}(21,-14)--(bra.215);
\node at (14.5,-21) {$\langle\psi|O\otimes I|\psi\rangle$};
\end{scope}
\begin{scope}[xshift=90mm]
\node[heading] at (0,5) {(d) Reduced density matrix};
\node[tensor] (psi) at (4,-8) {$\psi$};
\node[tensor] (bra) at (25,-8) {$\psi^*$};
\draw[wire] (psi.35)--++(0,5) node[label,above]{$i$};
\draw[wire] (bra.145)--++(0,5) node[label,above]{$i'$};
\draw[wire] (psi.325)--(8,-14)--node[label,below]{$j$}(21,-14)--(bra.215);
\node at (14.5,-21) {$\rho_{ii'}$};
\end{scope}
\end{tikzpicture}
\caption[Tensor-network constructions for common circuit outputs, illustrated for a two-qubit state. Here $\psi$ denotes the circuit tensor network and $\psi^*$ its complex conjugate. (a) Retain physical output indices to obtain a state tensor. (b) Attach computational-basis bras to all outputs to obtain a single amplitude. (c) Contract the ket and bra networks through an observable $O$ to obtain an expectation value. (d) Trace out qubit $j$ while retaining subsystem indices $i,i'$ to obtain a reduced density matrix.]{Tensor-network constructions for common circuit outputs, illustrated for a two-qubit state. Here $\psi$ denotes the circuit tensor network and $\psi^*$ its complex conjugate. (a) Retain physical output indices to obtain a state tensor. (b) Attach computational-basis bras to all outputs to obtain a single amplitude. (c) Contract the ket and bra networks through an observable $O$ to obtain an expectation value. (d) Trace out qubit $j$ while retaining subsystem indices $i,i'$ to obtain a reduced density matrix.}
\label{fig:quantum-output-networks}
\end{figure}
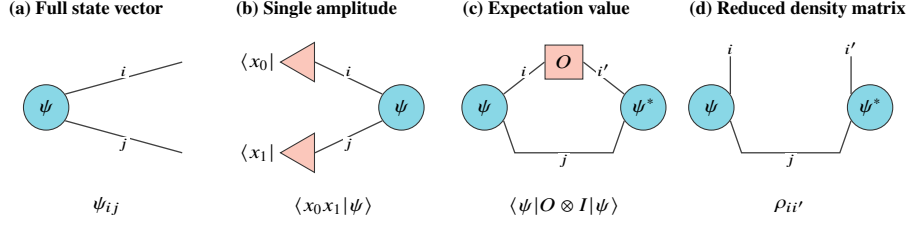

In quantum circuit simulation, local gates and initial states can be represented as tensors, with connected indices expressing the composition of operations~\cite{markov2008}, as in Figure~\ref{fig:intro-circuit}. The requested quantity determines the remaining connections and output labels. For example, fixing all final basis indices yields an amplitude, while retaining selected indices yields a tensor of amplitudes. Figure~\ref{fig:quantum-output-networks} shows these and other common constructions. Once this network has been constructed, contraction planning depends on its labels, dimensions, and requested output. Numerical tensor entries are needed only for execution.

\subsection{Contraction Orders and Their Costs}
\label{sec:orders}

For $n\geq2$, a binary contraction order $P$ consists of $n-1$ steps. At step $t$, two current tensors $A_t$ and $B_t$ are replaced by a result $C_t$. A label in either operand is retained if it occurs in another current tensor outside the pair or belongs to $O$. All other labels of the operands are summed over. This rule ensures that an index shared by several tensors is not eliminated prematurely. An order can be represented by a rooted binary tree whose leaves are input tensors and whose internal nodes are contraction results. A concrete execution order additionally specifies the sequence in which independent subtrees are evaluated.

Different orders compute the same expression in exact arithmetic, but their costs can differ considerably. Consider
\begin{equation}
  Y_{ad}=\sum_{b,c}A_{ab}B_{bc}C_{cd},
  \qquad
  A\in\mathbb C^{2\times8},\quad
  B\in\mathbb C^{8\times2},\quad
  C\in\mathbb C^{2\times8}.
  \label{eq:example}
\end{equation}
Contracting $A$ and $B$ first produces a $2\times2$ tensor, whereas contracting $B$ and $C$ first produces an $8\times8$ tensor. Table~\ref{tab:example} compares the resulting orders using the cost model below. Both produce the same $2\times8$ output, but the second order requires four times as many modeled operations and a fourfold larger maximum result tensor.

\begin{table}[htbp]
\centering
\small
\caption{Two orders for Eq.~\eqref{eq:example}. Step costs and total cost count modeled scalar multiplications, and $\Mmax$ counts elements, including the final output.}
\label{tab:example}
\begin{tabular}{@{}lccrr@{}}
\toprule
Order & First result & Step costs & $F(P)$ & $\Mmax(P)$ \\
\midrule
$(AB)C$ & $2\times2$ & $32+32$ & 64 & 16 \\
$A(BC)$ & $8\times8$ & $128+128$ & 256 & 64 \\
\bottomrule
\end{tabular}
\end{table}

Let $I(X)$ denote the set of distinct index labels of a tensor $X$, and let $|X|$ denote its number of elements. We define the modeled cost of a binary step and of a complete order as follows:\footnote{The count uses operand labels before elimination, including one-sided reductions, and excludes separate costs for unary preprocessing and backend-specific operations.}
\begin{equation}
  F_t=\prod_{e\in I(A_t)\cup I(B_t)}d_e,
  \qquad
  F(P)=\sum_{t=0}^{n-2}F_t.
  \label{eq:cost}
\end{equation}
Each distinct label contributes one dimension factor. Throughout this paper, \emph{FLOPs} refers to this scalar-multiplication count convention. This count is computed from the network structure and index dimensions, whereas the operations performed during execution depend on the backend.

Each result has one axis per retained label, so its size is $|C_t|=\prod_{e\in I(C_t)}d_e$, with an empty product equal to one. The largest contraction result is
\begin{equation}
  \Mmax(P)=\max_{0\leq t\leq n-2}|C_t|.
  \label{eq:size}
\end{equation}
This metric gives the number of elements in the largest individual contraction result, including the final output. To assess total process memory, we must also account for input storage, all simultaneously live results, element type, execution order, and backend workspace.

\subsection{Planning Objectives and Practical Tradeoffs}
\label{sec:planning}

Using the cost model above, the unsliced order optimization problem is
\begin{equation}
  \min_{P\in\mathcal P(\mathcal N)}F(P),
  \label{eq:optimization}
\end{equation}
where $\mathcal N$ denotes the network description, including input index-label tuples, index dimensions, and ordered output labels, and $\mathcal P(\mathcal N)$ denotes the set of all valid binary contraction orders for $\mathcal N$.

In practice, finding a low-cost order also takes time, so the planner must balance search effort against the expected execution savings. More extensive search may be worthwhile for an expensive contraction or when a plan is reused across many contractions with the same structure and shapes. For a small contraction, however, the search itself can dominate the total time. These differences motivate configurable search effort and stopping criteria that limit planning overhead.

Although $F(P)$ provides a common objective for comparing orders, execution time also depends on data layout, memory traffic, and the shapes of individual operations. We therefore evaluate both planning time and numerical execution time to assess the practical benefit of the returned orders. In addition, intermediate tensor sizes determine the storage needed to execute a plan, which motivates the constrained formulation below.

\subsection{Slicing under Tensor-Size Constraints}
\label{sec:slicing}

When a contraction result exceeds the available tensor-size budget, \emph{slicing} can reduce its size by fixing selected index variables and evaluating the resulting smaller contractions separately~\cite{gray2021}. Enumerating all assignments to the selected variables produces
\begin{equation}
  \Nslice=\prod_{e\in S}d_e
  \label{eq:slices}
\end{equation}
slice tasks, where $S\subseteq I$ denotes the set of sliced labels. Because each selected variable has a fixed value within a slice, its dimension is effectively one when evaluating that slice's cost and tensor sizes. The size requirement is then
\begin{equation}
  \Mmax(P\mid S)\leq M_{\mathrm{limit}},
  \label{eq:constraint}
\end{equation}
where $P\mid S$ denotes order $P$ evaluated with the sliced variables fixed, and $M_{\mathrm{limit}}$ is the maximum allowed number of elements in any result tensor within a slice, including its output.

For example, the order $A(BC)$ in Eq.~\eqref{eq:example} forms a 64-element intermediate. Slicing $b$ instead gives eight contractions indexed by $j=0,\ldots,7$:
\begin{equation}
  E_d^{(j)}=\sum_c B_{jc}C_{cd},
  \qquad
  Y_{ad}^{(j)}=A_{aj}E_d^{(j)},
  \qquad
  Y_{ad}=\sum_{j=0}^{7}Y_{ad}^{(j)}.
  \label{eq:sliceexample}
\end{equation}
Each $E^{(j)}$ has eight elements and each partial output has 16, so the sliced plan meets a limit of 16. However, its modeled total cost remains $8(16+16)=256$, excluding the final accumulation, whereas the unsliced order $(AB)C$ already meets the same limit at cost 64. This example shows why the contraction order and slice set should be considered together when seeking a low-cost feasible plan.

To compare such plans, we account for the work across all slices. Under a fixed dense schedule, every slice has the same modeled cost, giving
\begin{equation}
  \Fall(P,S)=\Nslice F(P\mid S).
  \label{eq:slicedcost}
\end{equation}
This expression assumes that slices are executed independently and excludes final accumulation and output assembly. Consequently, a contraction independent of a sliced variable may be repeated for each of its values, increasing the total work. We therefore seek an order and slice set that minimize this total while satisfying the size constraint:
\begin{equation}
  \min_{P\in\mathcal P(\mathcal N),\;S\subseteq E}
       \Fall(P,S)
  \quad\text{subject to}\quad
       \Mmax(P\mid S)\leq M_{\mathrm{limit}},
  \label{eq:constrained}
\end{equation}
where $E\subseteq I$ denotes the set of labels eligible for slicing. Each candidate plan is thus a pair $(P,S)$, evaluated by its total cost $\Fall(P,S)$ and its per-slice size $\Mmax(P\mid S)$.

To recover the final output, slice results are assembled according to the labels in $S$. When internal indices are sliced, the partial results are summed as in Eq.~\eqref{eq:sliceexample}, whereas slicing output indices produces distinct blocks of the final result. If both types are sliced, the partial results are summed within each output block. Although output slicing reduces the size of each block, materializing the complete output still requires storage for all its elements. This storage must be included when assessing total process memory, together with the other factors discussed in Section~\ref{sec:orders}.

\section{ArcTN: Architecture and Capabilities}
\label{sec:architecture}

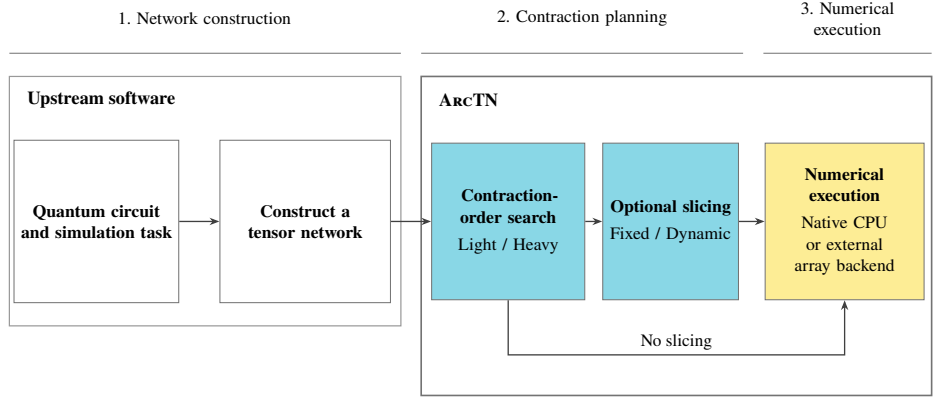
\begin{figure}[htbp]
\centering
\small
\definecolor{DiagramCyan}{HTML}{49C2D9}
\definecolor{DiagramOrange}{HTML}{F47254}
\definecolor{DiagramYellow}{HTML}{FDED95}
\resizebox{\linewidth}{!}{%
\begin{tikzpicture}[
  x=1mm,y=1mm,
  box/.style={draw=black!55,line width=0.5pt,fill=white,inner sep=1.2mm,align=center,font=\fontsize{8.5}{10.5}\selectfont},
  arrow/.style={-{Stealth[length=1.7mm]},draw=black!75,line width=0.65pt},
  values/.style={-{Stealth[length=1.7mm]},draw=DiagramOrange,dashed,line width=0.7pt},
  note/.style={font=\fontsize{8.5}{10.5}\selectfont,fill=white,inner sep=0.7mm}
]
\path[use as bounding box] (0,10) rectangle (159,81);
\filldraw[fill=white,draw=black!40,line width=0.6pt] (0,22) rectangle (67.5,65);
\node[anchor=west,font=\fontsize{9}{11}\selectfont\bfseries] at (2,61) {Upstream software};
\filldraw[fill=white,draw=black!55,line width=0.8pt] (71,10) rectangle (159,65);
\node[anchor=west,font=\fontsize{9}{11}\selectfont\bfseries,text=black] at (73,61) {\arctn{}};
\draw[black!55,line width=0.5pt] (0,69)--(67.5,69);
\draw[black!55,line width=0.5pt] (71,69)--(126.5,69);
\draw[black!55,line width=0.5pt] (130,69)--(159,69);
\node[align=center,font=\fontsize{8.5}{10.5}\selectfont,text width=66mm] at (33.75,75) {1. Network construction};
\node[align=center,font=\fontsize{8.5}{10.5}\selectfont,text width=55mm] at (98.75,75) {2. Contraction planning};
\node[align=center,font=\fontsize{8.5}{10.5}\selectfont,text width=29mm] at (144.5,75) {3. Numerical\\execution};
\node[box,text width=26mm,minimum height=28mm] (task) at (15,40) {\textbf{Quantum circuit and simulation task}};
\node[box,text width=27mm,minimum height=28mm] (network) at (51,40) {\textbf{Construct a}\\\textbf{tensor network}};
\node[box,fill=DiagramCyan!65,text width=24mm,minimum height=27mm] (search) at (86,40) {\textbf{Contraction-order search}\\[1mm] Light / Heavy};
\node[box,fill=DiagramCyan!65,text width=21mm,minimum height=27mm] (slice) at (114,40) {\textbf{Optional slicing}\\[1mm] Fixed / Dynamic};
\node[box,fill=DiagramYellow,text width=25mm,minimum height=27mm] (execute) at (144,40) {\textbf{Numerical execution}\\[1mm]
  Native CPU or external array backend};
\draw[arrow] (task.east)--(network.west);
\draw[arrow] (network.east)--(search.west);
\draw[arrow] (search.east)--(slice.west);
\draw[arrow] (slice.east)--(execute.west);
\draw[arrow] (search.south)--(86,17)--node[note,above,fill=white]{No slicing}(144,17)--(execute.south);
\end{tikzpicture}}
\caption[The ArcTN workflow. Upstream software constructs the tensor network from a circuit and a simulation task. ArcTN searches for a contraction order, optionally slices the network, and executes the contraction on its native CPU executor or an external array backend.]{The \arctn{} workflow. Upstream software constructs the tensor network from a circuit and a simulation task. \arctn{} searches for a contraction order, optionally slices the network, and executes the contraction on its native CPU executor or an external array backend.}
\label{fig:workflow}
\end{figure}

\arctn{} takes the network description from an upstream application and turns it into a contraction plan. Figure~\ref{fig:workflow} outlines this workflow. Planning uses the network structure and dimensions, while execution additionally consumes the numerical tensors. This separation allows the same planning interface to support direct contraction, integration with an external executor, and reuse with new tensor values.

\subsection{Planning Interfaces and Controls}
\label{sec:interfaces}

The Python interfaces expose the workflow at different levels. For applications that manage numerical execution themselves, \texttt{arctn\_path} returns a contraction path, while \texttt{arctn\_schedule} also reports its costs, sliced indices, and planning metadata. Applications that need an executable object can use \texttt{arctn\_plan}, and \texttt{arctn\_contract} provides a direct route to the numerical result. In addition, \texttt{ArcTNOptimizer} and \texttt{arctn\_tree} support integration through paths and Cotengra contraction trees.

Within these interfaces, \emph{Light} and \emph{Heavy} provide two levels of search effort. \emph{Light} emphasizes low planning overhead, whereas \emph{Heavy} devotes more work to candidate refinement. Users can also specify a random seed, a cooperative search-time limit, and a rate-based stopping condition. These controls allow the planning effort to reflect the expected cost and reuse of the contraction.

When a size target is requested, \emph{Fixed} or \emph{Dynamic} slicing augments the selected order with a slice set. The \texttt{target\_size} option specifies $M_{\mathrm{limit}}$ in Eq.~\eqref{eq:constraint}, and \texttt{allow\_output\_slicing} determines whether output labels are eligible. This option defaults to \texttt{False}. The direct execution interface supports both internal- and output-index slicing and assembles the complete result according to Section~\ref{sec:slicing}.

\subsection{Numerical Execution and Plan Reuse}
\label{sec:execution}

 The native CPU executor follows the contraction path and expresses each binary contraction as batched matrix multiplication. Shared labels retained in the current contraction result form batch dimensions, while eliminated shared labels form the matrix multiplication's inner dimension. After any one-sided reductions, axis permutations and reshapes expose these groups, and the executor calls the real or complex kernels of \texttt{matrixmultiply}. Alternatively, a fixed expression prepared by \texttt{opt\_einsum}~\cite{smith2018} delegates execution to an explicitly selected array backend, such as NumPy, CuPy, PyTorch, or JAX.

For repeated computations, \texttt{ArcTNExecutionPlan} stores the network structure, contraction path, and optional sliced indices. Because this object is serializable, a saved plan can be loaded and executed without repeating the search. For reuse within a running process, \texttt{ArcTNCompiledContraction} provides a compiled object for a fixed network and contraction path. This interface currently supports unsliced contractions and keeps its compiled state within the process. These two forms of reuse accommodate applications that either persist plans between runs or repeatedly execute one structure during a calculation.

\subsection{Availability}
\label{sec:availability}

The repository and documentation are available at \url{https://github.com/Quill-ArcLight/ArcTN}. %
The public repository also contains all
benchmark scripts, input manifests, and recorded results so that returned paths and reported costs can be inspected independently.

\section{Contraction Planning in ArcTN}
\label{sec:methods}

\arctn{} first constructs several candidate orders, then improves selected candidates and returns the lowest-cost order found. Keeping multiple starting points allows refinement to explore orders whose initial costs may not reflect their eventual quality. The framework combines random-greedy construction~\cite{gray2021}, subtree reconfiguration~\cite{cotengra082}, and parallel tempering~\cite{swendsen1986,hukushima1996}, with separate controls for the work spent at each stage.

\subsection{Candidate Generation}
\label{sec:generation}

Both \emph{Light} and \emph{Heavy} generate multiple initial contraction orders using random-greedy construction~\cite{gray2021}. To guide later attempts, both use the cross-entropy method (CEM) to adjust how construction parameters are sampled based on the costs of previously generated orders~\cite{rubinstein1999}. \emph{Light} additionally generates orders through recursive partitioning~\cite{gray2021}.

To obtain different starting points, both presets search the network supplied by the user (the original network) and a version produced by preprocessing (the simplified network). An order found for the simplified network is converted into a complete sequence of contractions for the original input tensors. All resulting orders can therefore be compared using the same cost $F(P)$ on the original network, and the most promising ones are selected for refinement. \emph{Heavy} also keeps the orders found on the simplified network as starting points for an additional search, described next.

Random-greedy construction builds an order one contraction at a time. At each step, it assigns a local heuristic cost to each pair of current tensors to guide the choice of the next contraction, with lower values preferred. One such cost is
\begin{equation}
  c_0(A,B)=|C|-\alpha\bigl(|A|+|B|\bigr),
  \label{eq:greedycost}
\end{equation}
where $C$ is the result of combining $A$ and $B$, $|\cdot|$ counts tensor elements, and $\alpha$ controls the weight given to the input sizes. This local heuristic cost favors a smaller result relative to its inputs. To generate different orders across attempts, the planner randomly chooses the next pair from at most eight pairs that share an index and have the lowest local heuristic costs, using
\begin{equation}
  \Pr(q)=\frac{\exp[-(c_q-c_{\min})/T]}{\sum_{r\in\mathcal B}\exp[-(c_r-c_{\min})/T]},
  \qquad T=\tau\max\{|c_{\min}|,1\},
  \label{eq:sampling}
\end{equation}
where $\mathcal B$ denotes the set of pairs considered, $c_q$ is the local heuristic cost of pair $q$, and $c_{\min}$ is the lowest such cost in that set. The parameter $\tau\geq 0$ controls randomness. For $\tau>0$,
pairs are sampled according to Eq.~\eqref{eq:sampling}, with
smaller values favoring lower-cost pairs more strongly.
For $\tau=0$, the planner follows the greedy rule and selects
a pair with the minimum local heuristic cost, breaking ties
arbitrarily. If the remaining tensors share no indices, it combines them by outer products, starting with the smallest tensors. Once the order is complete, its cost is evaluated using $F(P)$ from Eq.~\eqref{eq:cost}.

\subsection{Selective Refinement}
\label{sec:refinement}

After generating initial orders, the planner selects low-cost orders for improvement. It first changes small parts of each selected contraction tree, a procedure called \emph{subtree reconfiguration}~\cite{cotengra082}. The input tensors and output indices of the selected part remain fixed, while dynamic programming (DP) compares alternative ways to combine those inputs by reusing solutions to smaller subproblems. The replacement is accepted only if it lowers $F(P)$. \emph{Heavy} applies this refinement to more candidates than \emph{Light}, as summarized in Table~\ref{tab:presets}.

Local changes can stop improving an order even when better orders exist elsewhere in the search space. To explore further, the planner uses the best candidates to start an initial \emph{parallel-tempering} search~\cite{swendsen1986,hukushima1996}. This method runs several search sequences, or chains, at different temperatures. Temperature controls the willingness to accept a temporary cost increase, and chains can exchange their current orders. This allows the search to explore alternatives beyond those reached by local improvement alone. To decide whether further exploration is worthwhile, the planner checks how much this initial search has reduced the best cost from its value at the start of the search. If the reduction meets the requirement set by the chosen preset, it continues the same search to seek further improvements, keeping the lowest-cost order found throughout.

\emph{Heavy} also searches the simplified network through parallel tempering, followed by optional \emph{fixed-leaf-order dynamic programming}. This procedure keeps the input tensors in the same left-to-right order and searches for a lower-cost way to combine them in pairs. For example, it can change $((AB)C)D$ to $(AB)(CD)$ while preserving the sequence $A,B,C,D$. The resulting candidates are converted into orders for the original input tensors and compared using $F(P)$, after which the better order can receive a final subtree reconfiguration. Table~\ref{tab:presets} summarizes the refinement stages, and Algorithm~\ref{alg:scheduler} gives their execution conditions.

\begin{table}[htbp]
\centering
\small
\caption{Search structure of the two presets. Algorithm~\ref{alg:scheduler} specifies the checks that control entry to each stage.}
\label{tab:presets}
\begin{tabular}{@{}>{\raggedright\arraybackslash}p{0.28\linewidth}@{\hspace{14pt}}>{\raggedright\arraybackslash}p{0.29\linewidth}@{\hspace{14pt}}>{\raggedright\arraybackslash}p{0.34\linewidth}@{}}
\toprule
Stage & \emph{Light} & \emph{Heavy} \\
\midrule
Initial orders & Random-greedy with CEM and recursive partitioning & Random-greedy with CEM \\
Selected subtree refinement & Up to 2 candidates & Up to 8 candidates \\
Parallel tempering on the original network & Initial search with conditional continuation & Initial search with conditional continuation \\
Additional refinement & --- & Simplified-network tempering, optional fixed-leaf-order DP, final subtree reconfiguration \\
\bottomrule
\end{tabular}
\end{table}

\begin{algorithm}[htbp]
\caption{High-level scheduling of \emph{Light} and \emph{Heavy} order search}
\label{alg:scheduler}
\small
\begin{algorithmic}[1]
\Require Network $\mathcal N$, preset $p\in\{\text{\emph{Light}},\text{\emph{Heavy}}\}$
\Ensure Lowest-cost complete order $P^\star$ found on $\mathcal N$
\Statex \textit{Checks:} \Call{TimeRemains}{} is true before the time limit. \Call{SearchAllowed}{} additionally
\Statex \hspace{\algorithmicindent}checks the optional rate rule described in Section~\ref{sec:control}.
\State Generate candidates under $p$ on $\mathcal N$ and its simplified version $\widetilde{\mathcal N}$
\State $\mathcal C\gets$ distinct complete candidates converted to $\mathcal N$ and evaluated by $F$
\State $\mathcal Q\gets$ distinct initial orders on $\widetilde{\mathcal N}$
\If{$\mathcal C=\emptyset$}
  \State Construct and evaluate a greedy order on $\mathcal N$, then add it to $\mathcal C$
\EndIf
\State $\mathcal R\gets$ up to two (\emph{Light}) or eight (\emph{Heavy}) lowest-cost orders from $\mathcal C$
\While{$\mathcal R\neq\emptyset$ and \Call{SearchAllowed}{}}
  \State Remove an order $P$ from $\mathcal R$
  \State Apply subtree reconfiguration to $P$ and add the returned order to $\mathcal C$
\EndWhile
\If{$|\mathcal C|\geq2$ and \Call{SearchAllowed}{}}
  \State Run an initial parallel-tempering search from the two best orders in $\mathcal C$, adding its best order to $\mathcal C$
  \If{the initial search reduces the best cost sufficiently under $p$ and \Call{SearchAllowed}{}}
    \State Continue the same parallel-tempering chains and add their best order to $\mathcal C$
  \EndIf
\EndIf
\If{$p=\text{\emph{Heavy}}$, $|\mathcal Q|\geq2$, and \Call{SearchAllowed}{}}
  \State $Q\gets$ best order from parallel tempering on $\widetilde{\mathcal N}$ using candidates selected under $p$ from $\mathcal Q$
  \State Convert $Q$ to $P$ on $\mathcal N$ and evaluate $F(P)$
  \If{\Call{TimeRemains}{} and $\widetilde{\mathcal N}$ has at most $5000$ tensors}
    \State Apply fixed-leaf-order dynamic programming to $Q$
    \If{a complete order is returned}
      \State Convert the result to $P'$ on $\mathcal N$ and evaluate $F(P')$
      \If{$F(P')<F(P)$}
        \State $P\gets P'$
      \EndIf
    \EndIf
  \EndIf
  \If{\Call{TimeRemains}{}}
    \State $P\gets$ order returned by subtree reconfiguration of $P$
  \EndIf
  \State Add $P$ to $\mathcal C$
\EndIf
\State \Return $P^\star=\arg\min_{P\in\mathcal C}F(P)$
\end{algorithmic}
\end{algorithm}

\subsection{Search Control}
\label{sec:control}

The planner controls search effort at two levels. Each refinement stage stops when it no longer makes sufficient progress, allowing a later stage to try a different way of improving the order. The overall search is also bounded by a user-specified time limit and an optional rate rule that relates planning effort to the expected execution cost.

Following Cotengra~\cite{cotengra082}, the rate rule estimates execution time as $F(P_{\mathrm{best}})/r$, where $P_{\mathrm{best}}$ denotes the current best order and $r$ is an assumed number of modeled operations per second. Once elapsed search time exceeds this estimate, the rule prevents further work at the checks shown in Algorithm~\ref{alg:scheduler}. Thus, finding a cheaper order can also shorten the remaining search. Section~\ref{sec:setup} gives the rate values and time limits used in the experiments.

Algorithm~\ref{alg:scheduler} schedules candidate generation and refinement as described in Sections~\ref{sec:generation} and~\ref{sec:refinement}. Here, a complete order contracts all inputs to the requested output, and ``best'' means lowest modeled cost on the network being searched. All orders added to $\mathcal C$ are complete and evaluated using $F$ on the original network. For the candidate sets $\mathcal C$ and $\mathcal Q$, $|\cdot|$ denotes the number of orders.

\subsection{\emph{Fixed} and \emph{Dynamic} Slicing}
\label{sec:slicingmethods}

After choosing a contraction order, the planner can apply slicing to meet the tensor-size limit from Section~\ref{sec:problem}. The \emph{Fixed} slicing mode keeps that order unchanged and starts with no sliced labels. It progressively selects labels from results that exceed $M_{\mathrm{limit}}$, aiming to bring all results within the limit. The Boolean option \texttt{allow\_output\_slicing} controls whether labels of the final output may be selected for slicing.

Because slicing changes tensor sizes, an order chosen before slicing may no longer be a good choice for each slice. \emph{Dynamic} therefore alternates slice selection with local changes to the order. It first applies the \emph{Fixed} procedure to the given contraction order \(P\), selecting a slice set \(S\) such that \((P,S)\) meets the size limit. It then represents one slice by setting the selected index dimensions to one and applies subtree reconfiguration. A revised order $P'$ is accepted when it does not increase the cost of that slice, that is, $F(P'\mid S)\leq F(P\mid S)$. With this revised order, the next round selects labels again from an empty slice set and checks the size limit. Since a new slice set can change the number of tasks, \emph{Dynamic} compares the resulting feasible plans by their total cost $\Fall(P,S)$ and retains the cheapest one.

The two modes therefore offer different levels of planning effort. \emph{Fixed} selects slices for an existing order, while \emph{Dynamic} also revises the order to reduce the total sliced cost. Section~\ref{sec:slicingresults} compares them from the same initial orders and reports the settings used in the experiments.

\section{Implementation Design}
\label{sec:implementation}
This section describes how \arctn{} evaluates candidate costs in the dynamic programs of Section~\ref{sec:refinement}.

\paragraph{Exact integer costs.}
Subtree reconfiguration and fixed-leaf-order DP compare many candidate pairwise contractions. Because costs grow exponentially with the number of indices, a floating-point implementation compares them as base-2 logarithms, and adding two costs then requires a logarithm and an exponential. When every index has dimension two, as in all quantum-circuit networks evaluated here, \arctn{} instead represents these costs as exact 128-bit unsigned integers. A tensor with $k$ indices then has $2^k$ elements, obtained by a bit shift, and costs are combined by integer addition. Integer costs also compare tied orders exactly instead of up to rounding. In our tests, integer costs make the subtree-reconfiguration DP about 7\% faster and fixed-leaf-order DP 13--19\% faster without changing the results.

\paragraph{Overflow guard and fallback.}
Before using integer costs, \arctn{} checks a conservative bound on the cost of every complete order, $(n-1)\,2^{M}$ for the FLOPs objective, where $n$ is the number of input tensors and $M$ is the number of distinct indices or the largest number of axes of any input tensor, whichever is greater. Integer costs are used only if this bound fits in 128 bits, so they cannot overflow. Otherwise, or if some index has another dimension, the dynamic programs use the logarithmic arithmetic, so every network remains supported.

\section{Evaluation}
\label{sec:evaluation}

We evaluate how quickly \arctn{} finds low-cost orders, whether those orders reduce numerical execution time, how effectively slicing meets tensor-size limits, and whether these results carry over to accelerator hardware and to a larger collection of circuits. The experiments compare planning performance, execution through a shared numerical backend, and the costs of sliced plans.

\subsection{Experimental Setup}
\label{sec:setup}

\paragraph{Benchmarks.}
We use QC32, the 32 quantum-computing instances in Einsum Benchmark~\cite{einsumbenchmark}, comprising 20 scalar-output and 12 tensor-output networks (Appendix~\ref{sec:qc32-catalog}). These include random circuits, circuits for combinatorial optimization, and quantum Fourier transforms. All methods receive the same network descriptions, with every index having dimension two.

\paragraph{Baselines and configurations.}
We compare \emph{Light} and \emph{Heavy} with six baseline configurations from three tools. For Cotengra~\cite{gray2021}, we test Auto with CMA-ES~\cite{hansen2001} and AutoHQ with Optuna~\cite{akiba2019} for search-parameter tuning, along with Hyper128, which combines greedy construction, KaHyPar~\cite{schlag2022} network partitioning, and subtree reconfiguration. OMECO's TreeSA~\cite{omecorust,liu2026omeinsum,kalachev2021} improves contraction trees through simulated annealing (SA)~\cite{kirkpatrick1983}. Google TNCO~\cite{tnco} also uses simulated annealing, and we test it with and without tensor fusion. These configurations cover different ways of constructing and refining contraction orders.

Table~\ref{tab:searchsettings} gives the search budgets and stopping rules for all eight configurations. All target low FLOPs and are run five times per instance, with \emph{Light} and \emph{Heavy} using the rate rule from Section~\ref{sec:control}. We compare \emph{Light}, which limits planning overhead, with Cotengra Auto, the other low-overhead preset, which uses the same rate value. We compare \emph{Heavy}, which spends more effort on refinement, with AutoHQ, which uses the same rate value, and with Hyper128, OMECO, and TNCO, which search for low-cost orders until fixed trial or step limits. Figure~\ref{fig:planning-overview} shows all eight configurations together.

\begin{table}[!ht]
\centering
\caption{Main contraction-order search configurations. The planners check the listed time limits internally. Rate values are modeled operations per second.}
\label{tab:searchsettings}
\setlength{\tabcolsep}{8pt}
\renewcommand{\arraystretch}{1.08}
\begin{tabular}{@{}>{\raggedright\arraybackslash}p{0.27\linewidth}@{\hspace{14pt}}>{\raggedright\arraybackslash}p{0.34\linewidth}@{\hspace{14pt}}>{\raggedright\arraybackslash}p{0.30\linewidth}@{}}
\toprule
Configuration & Search settings & Limits and stopping \\
\midrule
\arctn{} \emph{Light} & \emph{Light} preset & 3600 s, rate $10^9$ \\
\arctn{} \emph{Heavy} & \emph{Heavy} preset & 3600 s, rate $10^8$ \\
Cotengra Auto/CMA-ES & Auto preset with CMA-ES & Up to 128 trials, rate $10^9$ \\
Cotengra AutoHQ/Optuna & AutoHQ preset with Optuna & Up to 128 trials, rate $10^8$ \\
Cotengra Hyper128/Optuna & Greedy and KaHyPar with Optuna, subtree size 12 & Up to 128 trials, 3500 s \\
OMECO TreeSA-8 & Simulated annealing & 8 trials, 50 iterations \\
Google TNCO SA & No fusion or decomposition of indices shared by more than two tensors & 8 trials, 1000 steps \\
Google TNCO SA (\texttt{fuse=4}) & Fusion enabled with the same SA settings & 8 trials, 1000 steps \\
\bottomrule
\end{tabular}
\end{table}

\paragraph{Environment and timing.}
Experiments ran on a server equipped with Intel Xeon Platinum 8253 processors at 2.20 GHz and approximately 2.95 TiB of memory. Every configuration used a fixed set of eight physical cores per run and a 3600-s external time limit. We ran up to 12 experiments concurrently on the server, with a limit of eight for TNCO. Table~\ref{tab:versions} lists the software versions.

\begin{table}[!ht]
\centering
\small
\caption{Software versions used in the evaluation. quimb constructs the MQT Bench networks of Section~\ref{sec:realcircuits}.}
\label{tab:versions}
\begin{tabular}{@{}llll@{}}
\toprule
Software & Version & Software & Version \\
\midrule
Python & 3.12.3 & NumPy & 2.2.6 \\
\arctn{} & 1.0.0 & \texttt{opt\_einsum} & 3.4.0 \\
Cotengra & 0.8.2 & Optuna & 4.9.0 \\
\texttt{cotengrust} & 0.2.1 & CMA-ES package & 0.13.0 \\
OMECO & 0.2.6 & KaHyPar & 1.3.5 \\
Google TNCO & 0.4.0 & OpenBLAS & 0.3.29 \\
quimb & 1.15.0 & & \\
\bottomrule
\end{tabular}
\end{table}
Planning time is measured from calling the planner until it returns, excluding numerical execution.\footnote{For TNCO, later reconstruction and selection of orders are excluded from this time. Tensor fusion is included in both planning time and reported cost.} We evaluate all returned plans on the original networks using the same cost model from Section~\ref{sec:problem}.

\paragraph{Metrics and aggregation.}
Besides FLOPs $F(P)$ and largest-result size $\Mmax(P)$, we report total output writes $W(P)=\sum_t|C_t|$, the total number of result elements produced across all contraction steps. Peak live elements $L_{\max}(P)$ measure the maximum combined size of input tensors and contraction results stored at the same time, in elements.

In the planning experiment (Section~\ref{sec:planningresults}), the median of five runs gives a representative value for each instance. We then take the geometric mean of these values across instances for which every configuration completed all five runs. Speedup is the baseline planning time divided by \arctn{}'s planning time. In the execution experiment (Section~\ref{sec:executionresults}), we take geometric means over runs for the same instances and repeats that all compared configurations completed. In the slicing experiments (Section~\ref{sec:slicingresults}), we first take the geometric mean over five runs per instance and then across instances. Section~\ref{sec:realcircuits} defines the ratios reported for MQT Bench.

\subsection{Planning Efficiency and Contraction Quality}
\label{sec:planningresults}

OMECO completed 152 of 160 planning runs, while every other compared configuration completed all 160. To compare all configurations on the same instances, we exclude the two instances for which OMECO did not complete all five runs. The comparison therefore covers the remaining 30 instances, each with five successful runs per configuration.

\begin{figure}[htbp]
\centering
\includegraphics[width=0.9\linewidth]{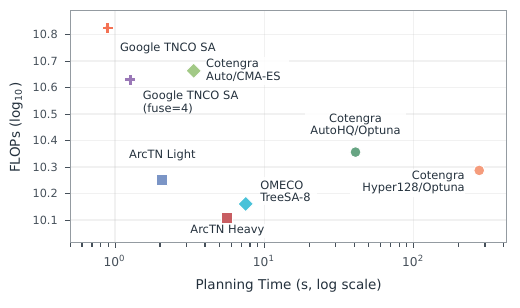} \caption{Planning time (horizontal axis) and FLOPs (vertical axis) averaged over all 30 instances. Points in the bottom left indicate better performance.
} \label{fig:planning-overview}
\end{figure}

Figure~\ref{fig:planning-overview} summarizes planning time and contraction cost on these 30 instances. \emph{Light} spends less time planning than \emph{Heavy}, while \emph{Heavy} achieves the lowest modeled FLOPs among all eight configurations. \emph{Light} improves on all three Cotengra configurations in both planning time and FLOPs, and \emph{Heavy} improves on AutoHQ and Hyper128 in both. Table~\ref{tab:planningresults} quantifies these benefits. \emph{Light} is $1.62\times$ faster than Cotengra Auto and reduces FLOPs by 61.32\%, while \emph{Heavy} is $7.21\times$ faster than Cotengra AutoHQ with 43.83\% fewer FLOPs and also outperforms OMECO on both measures. Compared with TNCO, \emph{Heavy} spends longer searching and obtains substantially lower FLOPs.

\begin{table}[htbp]
\centering
\small
\caption{Planning comparisons on the same 30 instances, with five completed runs per configuration on each instance. Reductions are percentages relative to the baseline. Writes, largest, and peak denote $W(P)$, $\Mmax(P)$, and $L_{\max}(P)$.}
\label{tab:planningresults}
\begin{tabular}{@{}lrrrrr@{}}
\toprule
Comparison & Search & \multicolumn{4}{c}{Reduction (\%)} \\
\cmidrule(l){3-6}
\arctn{} / baseline & speedup & FLOPs & Writes & Largest & Peak \\
\midrule
\emph{Light} / Auto & $1.62\times$ & 61.32 & 64.16 & 59.39 & 61.30 \\
\emph{Heavy} / AutoHQ & $7.21\times$ & 43.83 & 34.96 & 39.85 & 38.81 \\
\emph{Heavy} / Hyper128 & $48.64\times$ & 34.15 & 37.04 & 42.57 & 42.50 \\
\emph{Heavy} / OMECO & $1.33\times$ & 11.85 & 6.89 & 16.88 & 18.72 \\
\emph{Heavy} / TNCO & $0.16\times$ & 80.93 & 35.86 & 80.16 & 78.31 \\
\emph{Heavy} / TNCO (\texttt{fuse=4}) & $0.22\times$ & 70.07 & 21.51 & 67.01 & 68.38 \\
\bottomrule
\end{tabular}
\end{table}

\emph{Heavy} has the lowest geometric mean for all four reported cost and storage measures among the compared configurations. However, a lower modeled cost does not directly establish an execution-time benefit, so we next execute the returned orders using the same numerical backend.

\subsection{Numerical Execution Benefits}
\label{sec:executionresults}

We execute unsliced orders from the five configurations with the lowest aggregate FLOPs, using \texttt{opt\_einsum} with NumPy and the same saved input arrays. The inputs comprise 20 \texttt{complex128} and 12 \texttt{complex64} instances. Each run uses eight physical cores and eight OpenBLAS threads for one measured contraction without warm-up.

For each configuration, all 32 instances are scheduled to run five times, resulting in 160 runs per configuration. Across the five configurations, 747 of the 800 runs completed. The others exceeded storage limits,\footnote{We skip a run if any contraction result is expected to require more than 64 GiB, or the final output more than 32 GiB. The limit is 128 GiB for all tensors stored at once, and also for the two inputs and result of any single contraction combined.} lacked a saved order, or reached the one-hour limit. All five configurations successfully completed the same 134 runs across 29 instances. We use only these runs to calculate the geometric means of FLOPs and execution times in Table~\ref{tab:executionresults}.

Table~\ref{tab:executionresults} reports contraction time (Contraction column) and the sum of the times for the planning and execution stages, including their overhead (Total column). \emph{Heavy} achieves the lowest geometric mean for both measures, reducing total time by 13.99\% relative to \emph{Light}, 21.97\% relative to OMECO, and 53.67\% relative to AutoHQ. Thus, its execution benefits also translate into lower total time when planning is included.

\begin{table}[htbp]
\centering
\small
\caption{Numerical execution results. Completed runs are counted out of 160 planned runs per configuration. FLOPs and times are geometric means over the 134 runs completed by all five configurations. The last column gives the reduction in Total achieved by \emph{Heavy} relative to each row.}
\label{tab:executionresults}
\begin{tabular}{@{}lrrrrr@{}}
\toprule
Configuration & Completed & FLOPs & Contraction & Total & \emph{Heavy}'s \\
 & runs & ($10^9$) & (s) & (s) & reduction (\%) \\
\midrule
\arctn{} \emph{Heavy} & 155/160 & 9.425 & 43.98 & 67.80 & --- \\
\arctn{} \emph{Light} & 150/160 & 12.394 & 61.92 & 78.83 & 13.99 \\
OMECO TreeSA-8 & 148/160 & 9.885 & 50.84 & 86.88 & 21.97 \\
Cotengra AutoHQ & 146/160 & 16.175 & 60.74 & 146.34 & 53.67 \\
Cotengra Hyper128 & 148/160 & 13.536 & 57.55 & 408.32 & 83.40 \\
\bottomrule
\end{tabular}
\end{table}

\subsection{Planning under Tensor-Size Constraints}
\label{sec:slicingresults}

The preceding experiments use unsliced orders. We now require every contraction result to fit a size limit. We compare four planning configurations on 18 scalar-output and all 12 tensor-output QC32 instances, with five runs per configuration on each instance. The scalar set contains the 18 of 20 instances for which every configuration completed all five runs. Table~\ref{tab:slicesettings} lists the settings. \emph{Fixed} and \emph{Dynamic} start from the same saved \emph{Heavy} order for each instance and repeat.

\begin{table}[htbp]
\centering
\small
\caption{Order-search and slicing configurations.}
\label{tab:slicesettings}
\begin{tabular}{@{}>{\raggedright\arraybackslash}p{0.24\linewidth}@{\hspace{14pt}}>{\raggedright\arraybackslash}p{0.31\linewidth}@{\hspace{14pt}}>{\raggedright\arraybackslash}p{0.36\linewidth}@{}}
\toprule
Configuration & Initial order & Slicing stage \\
\midrule
\arctn{} \emph{Fixed} & Saved \emph{Heavy} path & Select indices to slice; keep the order unchanged \\
\arctn{} \emph{Dynamic} & Same saved \emph{Heavy} path & 3 rounds, size-8 subtree reconfiguration \\
Cotengra AutoHQ & Optuna, FLOPs objective, 128 trials, 8 workers & Slice the returned tree with its order fixed \\
OMECO & TreeSA with 8 trials and 50 iterations & TreeSASlicer with 8 trials, 10 iterations, optimization ratio 2.0, and $\texttt{sc\_target}=\log_2 M_{\mathrm{limit}}$ \\
\bottomrule
\end{tabular}
\end{table}

The size limit is $2^{26}$ elements per contraction result for scalar outputs and $\min(2^{26},\lfloor|Y_c|/4\rfloor)$ for tensor outputs, where $|Y_c|$ is the output size. The tensor-output experiment allows all four configurations to slice output indices. The two groups use different size targets and are reported separately.

Table~\ref{tab:crossslicing} reports total sliced FLOPs and full planning time, including order search and slicing. Values are geometric means over five runs within each instance and then across instances. These experiments generate and validate plans without executing numerical sliced contractions. FLOPs exclude final accumulation and output assembly.

\begin{table}[htbp]
\centering
\small
\caption{Slicing results for the evaluated configurations. Percentage changes are for \emph{Dynamic} relative to the configuration in each row. Negative values indicate lower FLOPs or shorter full planning time.}
\label{tab:crossslicing}
\begin{tabular}{@{}llrr@{}}
\toprule
Instances & Configuration & FLOPs change (\%) & Time change (\%) \\
\midrule
18 scalar outputs & \arctn{} \emph{Fixed} & $-31.99$ & $+17.40$ \\
 & OMECO & $+1.06$ & $-29.51$ \\
 & Cotengra AutoHQ & $-92.43$ & $-94.26$ \\
\addlinespace
12 tensor outputs & \arctn{} \emph{Fixed} & $-27.04$ & $+17.25$ \\
 & OMECO & $-49.71$ & $+11.08$ \\
 & Cotengra AutoHQ & $-71.18$ & $-82.21$ \\
\bottomrule
\end{tabular}
\end{table}

\Needspace{6\baselineskip} All retained plans meet their size limits. \emph{Dynamic} reduces total FLOPs relative to \emph{Fixed} by 31.99\% for scalar outputs and 27.04\% for tensor outputs, with additional planning time. Compared with OMECO, its scalar-output FLOPs are 1.06\% higher with 29.51\% less planning time, and its tensor-output FLOPs are 49.71\% lower with 11.08\% more planning time. The last two experiments vary the hardware and the circuits. Section~\ref{sec:hygon} executes the returned orders on accelerator hardware, and Section~\ref{sec:realcircuits} compares the planners on a larger collection of circuits.

\subsection[Deployment and Evaluation on the Hygon Cluster]{Deployment and Evaluation on the Hygon Cluster}
\label{sec:hygon}

To assess \arctn{} on accelerator hardware, we compare \emph{Light} and \emph{Heavy} with Cotengra\footnote{We only compare against Cotengra here (rather than all configurations considered elsewhere) because the Hygon machine natively provides Cotengra as the tensor contraction planning stack.} on a server in the Hygon cluster, using QC32 and a tensor network library jointly developed by Arclight and Sugon on top of Sugon's tensor library. The server is equipped with two Hygon C86 7490 64-core CPUs (2.7~GHz) and eight Hygon BW200 Deep Computing Unit (DCU) accelerators, each with 64~GiB of device memory. All experiments use a single DCU. All configurations use \texttt{complex128} and seed 0. Cotengra's \texttt{HyperOptimizer} minimizes FLOPs with up to 16 trials. We report planning time for finding an order (Planning), numerical contraction time (Contraction), and total time including planning, execution, and associated overhead (Total). Total is measured once per circuit and configuration, while Contraction is measured separately as the median of five runs after two warm-ups.

Both \arctn{} presets completed the same 14 of 32 circuits. Cotengra was evaluated on these 14 and completed six, which we use for all performance comparisons. Table~\ref{tab:hygon} shows that both presets reduce planning and contraction time, achieving speedups in total time of $5.91\times$ for \emph{Light} and $6.53\times$ for \emph{Heavy}. Although \emph{Heavy} spends longer planning than \emph{Light}, it achieves the lowest contraction and total times on these six circuits.

\begin{table}[htbp]
\centering
\small
\caption{Hygon server results: geometric means over the six circuits completed by all three configurations. Contraction uses per-circuit medians. Logical FLOPs count two operations per multiply--add. Speedup is Cotengra's Total divided by each configuration's Total.}
\label{tab:hygon}
\begin{tabular}{@{}lrrr@{}}
\toprule
Metric & Cotengra & \arctn{} \emph{Light} & \arctn{} \emph{Heavy} \\
\midrule
Total (s) & 57.95 & 9.81 & 8.87 \\
Planning (s) & 33.19 & 1.10 & 3.30 \\
Contraction (s) & 1.716 & 0.670 & 0.563 \\
Logical FLOPs ($10^9$) & 13.829 & 4.141 & 3.073 \\
Speedup (Total) & $1.00\times$ & $5.91\times$ & $6.53\times$ \\
\bottomrule
\end{tabular}
\end{table}

\subsection{Evaluation on Additional Realistic Quantum Circuits}
\label{sec:realcircuits}

To test whether the QC32 results extend to other circuits and simulation tasks, we compare the same planners on tensor networks built from MQT Bench~\cite{mqtbench}. Appendix~\ref{app:mqt} gives further details and complete results.

\paragraph{Circuits.}
We take 320 circuits of MQT Bench v1.1.9 at the target-independent level. For the 14 families provided at every qubit count up to 130, such as amplitude estimation, phase estimation, and quantum neural networks, we take $n=10,20,\ldots,130$ (182 circuits). The other 14 families, such as QAOA, VQE, Grover search, and quantum walks, encode specific problem instances at a few sizes between 2 and 31 qubits, and we take all 138 of their circuits. MQT Bench fixes each circuit's depth by its algorithm and qubit count. Table~\ref{tab:mqtcircuits} lists the qubit counts, depths, and gate counts of each family.

\paragraph{Tasks and networks.}
For a circuit $C$ on $n$ qubits, we build four networks with the constructions of Figure~\ref{fig:quantum-output-networks}. The \emph{amplitude} network gives $\langle0^n|C|0^n\rangle$. The two \emph{expectation} networks give $\langle0^n|C^\dagger Z_cC|0^n\rangle$ and $\langle0^n|C^\dagger Z_cZ_{c+1}C|0^n\rangle$ with $c=\lfloor(n-1)/2\rfloor$. The \emph{state} network keeps all qubits open for $n\le28$, and otherwise keeps 20 evenly spaced qubits open and fixes the rest to $|0\rangle$. quimb~1.15~\cite{gray2018quimb} constructs each network and simplifies it, which can produce indices shared by more than two tensors. All planners, including TNCO, receive the same simplified network, and construction and simplification are excluded from planning time.

Table~\ref{tab:mqtnetworks} counts the networks by their number of tensors after simplification. Networks with at most two tensors need no search, 14 networks of the Grover and quantum-walk families without ancillas were not constructed within 3600~s, and we leave out the 197 networks with 5000 or more tensors. We plan the remaining 672 networks, 530 small ones with fewer than 1000 tensors and 142 large ones. The tasks keep different numbers of networks. Simplification often reduces an amplitude network, with both ends fixed to $|0^n\rangle$, to one or two tensors, whereas an expectation network contains gates from both $C$ and $C^\dagger$ and a state network keeps its open indices.

\begin{table}[htbp]
\centering
\small
\setlength{\tabcolsep}{6pt}
\caption{MQT Bench networks by task. Parentheses give the number of tensors after simplification, or the construction time for Timeout. Only small and large networks are planned.}
\label{tab:mqtnetworks}
\begin{tabular}{@{}lrrrrrr@{}}
\toprule
Task & Networks & Trivial & Small & Large & Very large & Timeout \\
 & & (1--2) & (3--999) & (1000--4999) & ($\ge5000$) & ($>3600$~s) \\
\midrule
Amplitude & 320 & 191 & 66 & 19 & 40 & 4 \\
$\langle Z_c\rangle$ & 320 & 101 & 113 & 47 & 56 & 3 \\
$\langle Z_cZ_{c+1}\rangle$ & 320 & 98 & 115 & 48 & 56 & 3 \\
State & 320 & 7 & 236 & 28 & 45 & 4 \\
\midrule
Total & 1280 & 397 & 530 & 142 & 197 & 14 \\
\bottomrule
\end{tabular}
\end{table}

\paragraph{Configurations.}
We use the configurations of Table~\ref{tab:searchsettings} with three changes. First, rate-based stopping is disabled, because on small networks it would stop the search almost immediately, at the cost of higher FLOPs. Each planner then stops at its trial or iteration limit, at the end of its schedule, or at its time limit (3600~s for \emph{Light} and \emph{Heavy}, 3500~s for Cotengra). Second, OMECO TreeSA-8 also runs with 200 iterations, because 50 iterations return much worse orders on large networks. Third, we report one run per configuration and network. For each metric, we divide a configuration's value on each network by that of \emph{Heavy} and report the geometric mean of these ratios across networks. All $9\times672$ runs returned a valid order.

\paragraph{Results.}
On the 530 small networks, \emph{Heavy}, AutoHQ, Hyper128, and OMECO with 200 iterations are within 4\% of each other in geometric-mean FLOPs, while \emph{Heavy} plans $1.26\times$, $61\times$, and $2.25\times$ faster than the other three (Tables~\ref{tab:mqtsummary} and~\ref{tab:mqtresults}). On the 142 large networks, \emph{Heavy} has the lowest geometric-mean FLOPs. AutoHQ and Hyper128 require $1.34\times$ and $1.23\times$ its FLOPs with $3.1\times$ and $31\times$ its planning time, and \emph{Light} requires $1.29\times$ its FLOPs in about half its time.

\begin{table}[htbp]
\centering
\footnotesize
\setlength{\tabcolsep}{3.2pt}
\caption{MQT Bench results relative to \emph{Heavy}. Entries are geometric means, over networks, of each configuration's FLOPs or planning time divided by that of \emph{Heavy}, so values above one favor \emph{Heavy}. Numbers of networks are given in parentheses, and expectation combines $\langle Z_c\rangle$ and $\langle Z_cZ_{c+1}\rangle$. Tables~\ref{tab:mqtresults} and~\ref{tab:mqttasks} in Appendix~\ref{app:mqt} add writes, largest result, and peak live elements.}
\label{tab:mqtsummary}
\begin{tabular}{@{}lrrrrrrr@{}}
\toprule
 & \multicolumn{6}{c}{FLOPs ratio} & Time ratio \\
\cmidrule(lr){2-7}\cmidrule(l){8-8}
 & All & Small & Large & Ampl. & Expect. & State & All \\
Configuration & (672) & (530) & (142) & (85) & (323) & (264) & (672) \\
\midrule
\arctn{} \emph{Heavy} & 1.00 & 1.00 & 1.00 & 1.00 & 1.00 & 1.00 & 1.00 \\
\arctn{} \emph{Light} & 1.12 & 1.07 & 1.29 & 1.17 & 1.13 & 1.09 & 0.39 \\
Cotengra Hyper128 & 1.04 & 0.99 & 1.23 & 1.72 & 0.80 & 1.22 & 52.7 \\
Cotengra AutoHQ & 1.07 & 1.01 & 1.34 & 1.82 & 0.84 & 1.22 & 1.53 \\
Cotengra Auto & 34.3 & 2.51 & $6.0\times10^{5}$ & 23.4 & 262 & 3.23 & 0.45 \\
OMECO TreeSA-8 (50 iter.) & 11.8 & 1.08 & $8.8\times10^{4}$ & 3.03 & 96.5 & 1.39 & 0.51 \\
OMECO TreeSA-8 (200 iter.) & 5.08 & 1.03 & $2.0\times10^{3}$ & 2.10 & 20.7 & 1.21 & 1.91 \\
TNCO & 207 & 3.97 & $5.3\times10^{8}$ & 10.7 & $3.3\times10^{3}$ & 18.3 & 0.33 \\
TNCO (\texttt{fuse=4}) & 42.6 & 2.69 & $1.3\times10^{6}$ & 6.74 & 414 & 4.76 & 0.33 \\
\bottomrule
\end{tabular}
\end{table}

By task, \emph{Heavy} has the lowest geometric-mean FLOPs on amplitude and state networks, whereas on expectation networks AutoHQ and Hyper128 reach $0.84\times$ and $0.80\times$ its FLOPs with $3.6\times$ and $61\times$ its planning time (Table~\ref{tab:mqttasks}). With 200 iterations, OMECO returns lower FLOPs than \emph{Heavy} on 96 of the 142 large networks, but its geometric mean on them is $2.0\times10^3$ times that of \emph{Heavy} ($8.8\times10^4$ with 50 iterations). Cotengra Auto and both TNCO configurations plan quickly but return orders with several orders of magnitude more FLOPs on the large networks.

\section{Related Tools}
\label{sec:related}

Existing contraction planners also combine initial order construction with later improvement. Cotengra constructs orders using greedy choices or partitions of the network, automatically tunes the construction parameters, and supports local tree changes, simulated annealing, and slicing~\cite{gray2021}. OMECO provides a Rust implementation of OMEinsumContractionOrders~\cite{liu2026omeinsum} with Python interfaces. Its TreeSA method~\cite{kalachev2021} improves contraction trees through simulated annealing, while TreeSASlicer selects slices to control tensor sizes~\cite{omecorust}. Google TNCO similarly provides Python interfaces to a native implementation and searches contraction trees through simulated annealing, with an option to combine tensors before searching~\cite{tnco}. We compare against these tools because they address the same practical tasks of finding contraction orders and managing their computational or storage costs.

Other libraries connect contraction planning to applications and execution. For example, quimb constructs tensor networks for quantum calculations and can ask external optimizers to find contraction orders~\cite{gray2018quimb}, while \texttt{opt\_einsum} prepares reusable contraction expressions for different numerical libraries~\cite{smith2018}. For execution on other hardware, cuTensorNet combines planning with GPU computation~\cite{cutensornetoverview}, and TNC distributes tensor contractions across processes using the Message Passing Interface (MPI)~\cite{geiger2026tnc}. \arctn{} supports these kinds of workflows by exposing orders as paths or trees and by providing executable plans that can use its own executor or an external numerical library. Its engineering contribution combines candidate generation, selected-order refinement, stopping rules, and slicing in one configurable tool.

\section{Conclusion}
\label{sec:conclusion}

We presented \arctn{}, a tool that integrates contraction-order search, slicing, and reusable execution plans through Python interfaces to a Rust implementation. Its \emph{Light} and \emph{Heavy} presets coordinate candidate generation and refinement with controls on search effort, while \emph{Fixed} and \emph{Dynamic} slicing support tensor-size constraints. Together, these capabilities connect planning and numerical execution in a configurable workflow for quantum circuit simulation.

On QC32, \arctn{} finds lower-cost orders in less planning time than several baseline configurations. When the returned orders are executed with the same NumPy backend, \emph{Heavy} achieves the lowest geometric-mean contraction time and total planning--execution time among the five tested configurations. \emph{Dynamic} reduces total sliced FLOPs relative to \emph{Fixed} for both scalar and tensor outputs, with additional planning effort, and both modes meet every requested size limit. On networks built from MQT Bench circuits, \emph{Heavy} also obtains the lowest geometric-mean FLOPs on the large networks, with much less planning time than Cotengra AutoHQ and Hyper128. These results demonstrate the practical benefits of coordinating order search and slicing to balance planning effort, contraction cost, and tensor size. Future work includes improving \emph{Heavy} on expectation-value networks and extending the native executor to accelerators.

\bibliographystyle{splncs04}
\bibliography{references}

\appendix
\renewcommand{\theHsection}{appendix.\Alph{section}}
\section{QC32 Circuit Set}
\label{sec:qc32-catalog}

Table~\ref{tab:qc32-catalog} lists all 32 QC32 networks, including the two MaxCut instances omitted from the common planning comparison set. Instance identifiers omit the common \texttt{qc\_} prefix.

\begin{table}[htbp]
\caption{Circuit types, sources, and output shapes in QC32}
\label{tab:qc32-catalog}
\centering
\footnotesize
\def\UrlBreaks{\do\_\do\-}
\setlength{\tabcolsep}{4pt}
\begin{tabular}{@{}>{\raggedright\arraybackslash}p{24mm}>{\raggedright\arraybackslash}p{76mm}>{\raggedright\arraybackslash}p{16mm}@{}}
\toprule
Type/source (count) & Instance identifiers (without \texttt{qc\_}) & Output (rank $r$) \\
\midrule
\multicolumn{3}{@{}l}{\textit{Google/Sycamore random circuits (9)}} \\
\addlinespace[2pt]
Named instances (6) &
\nolinkurl{circuit_n47_m14_s0_e0_pEFGH}\newline
\nolinkurl{circuit_n49_m14_s9_e6_pEFGH_simplified}\newline
\nolinkurl{circuit_n53_m12_s2_e8_pABCDCDAB_simplified}\newline
\nolinkurl{circuit_n53_m20_s5_e22_pABCDCDAB}\newline
\nolinkurl{circuit_patch_n53_m16_s10_e28_pABCDCDAB}\newline
\nolinkurl{circuit_patch_n53_m16_s19_e28_pABCDCDAB_simplified} & Scalar\newline $r=0$ \\
\addlinespace
Yao \texttt{rand\_}\texttt{google\allowbreak 53} (3) &
\nolinkurl{rand_google53_25_50}\newline
\nolinkurl{rand_google53_26_15}\newline
\nolinkurl{rand_google53_27_20} & Tensor\newline $r=25,$\newline $26,27$ \\
\midrule
Numbered benchmark circuits (8) &
\nolinkurl{circuits-1}, \nolinkurl{circuits-2},
\nolinkurl{circuits-7}, \nolinkurl{circuits-8},
\nolinkurl{circuits-14}, \nolinkurl{circuits-18},
\nolinkurl{circuits-35}, \nolinkurl{circuits-42} & Scalar\newline $r=0$ \\
\addlinespace
MaxCut QAOA (6) &
\nolinkurl{maxcut_n18_r17_p100}\newline
\nolinkurl{maxcut_n24_r20_p10}\newline
\nolinkurl{maxcut_n25_r4_p4}\newline
\nolinkurl{maxcut_n27_r6_p5}\newline
\nolinkurl{maxcut_n28_r3_p3}\newline
\nolinkurl{maxcut_n29_r4_p10} & Scalar\newline $r=0$ \\
\addlinespace
Quantum Fourier transform (3) &
\nolinkurl{qft_27}, \nolinkurl{qft_28}, \nolinkurl{qft_30} & Tensor\newline $r=27,$\newline $28,30$ \\
\addlinespace
Yao grid random circuits (3) &
\nolinkurl{rand_supremacy2d_1_12_400}\newline
\nolinkurl{rand_supremacy2d_4_4_100}\newline
\nolinkurl{rand_supremacy2d_7_2_800} & Tensor\newline $r=23,$\newline $28,21$ \\
\addlinespace
Yao variational circuits (3) &
\nolinkurl{variational_29}\newline
\nolinkurl{variational_30}\newline
\nolinkurl{variational_31} & Tensor\newline $r=29,$\newline $30,31$ \\
\midrule
Total (32) & \multicolumn{2}{l@{}}{20 scalar outputs; 12 tensor outputs} \\
\bottomrule
\end{tabular}
\par\smallskip
\raggedright
Rank $r$ is the number of open output indices. All indices have dimension 2, giving $2^r$ output elements. Rank lists follow the order of the instance identifiers within each row.

\end{table}

Two pairs of instances have identical input labels, dimensions, output labels, and tensor shapes, namely \nolinkurl{circuit_n53_m20_s5_e22_pABCDCDAB} and \nolinkurl{circuits-14}, and \nolinkurl{circuits-42} and \nolinkurl{circuits-7}. In each pair, the former uses \texttt{complex128} and the latter \texttt{complex64}, with different numerical arrays. The 32 instances therefore contain 30 distinct structural descriptions, and the 30-instance main comparison contains 28. We retain the archived instances and give each included instance equal weight in the planning aggregates.

The output ranks describe the saved contraction problems and do not uniquely identify the original physical tasks. In particular, the grid instances have more open output indices than qubits, so their tensor outputs should not be interpreted as full state vectors.

The numbered instances retain their benchmark identifiers because their original gate-level circuits are unavailable. For the named families, filenames provide additional information. MaxCut identifiers give the vertex count \texttt{n}, regular-graph degree \texttt{r}, and QAOA depth \texttt{p}. In \texttt{rand\_google53}, ``53'' belongs to the generator name, and the following numbers give the qubit count and depth. For grid instances, the first two numbers specify the layout and the last gives the depth.

\section{MQT Bench Benchmark Details}
\label{app:mqt}

This appendix gives the details of the evaluation in Section~\ref{sec:realcircuits}.

\subsection{Circuit selection}
We downloaded the complete pre-generated archive of MQT Bench v1.1.9 and selected circuits by file name only, before any planning run. We use the target-independent level, compiled with Qiskit~\cite{qiskitdocs}. The circuit files were generated with MQT Bench 1.1.0 and Qiskit 1.0.2. Every family of this level is included. MQT Bench provides 14 families at every qubit count from 2 (3 for graph states) to 130. They are amplitude estimation (ae), Deutsch--Jozsa (dj), GHZ-state, graph-state, and W-state preparation (ghz, graphstate, wstate), the quantum Fourier transform with and without entangled input (qft, qftentangled), quantum neural networks (qnn), exact and inexact phase estimation (qpeexact, qpeinexact), random circuits (random), and three variational ans\"atze with random parameters (realamprandom, su2random, twolocalrandom). For these, we take $n=10,20,\ldots,130$. The other 14 families encode specific problem instances and are provided only at a few sizes between 2 and 31 qubits, so we take all of their circuits. They comprise Grover search and quantum walks, each with and without ancilla qubits (grover-noancilla, grover-v-chain, qwalk-noancilla, qwalk-v-chain), QAOA and VQE (qaoa, vqe), portfolio optimization with QAOA or VQE (portfolioqaoa, portfoliovqe), option pricing for call and put options (pricingcall, pricingput), the traveling salesperson problem (tsp), a routing problem (routing), Shor's algorithm with two instances (shor), and ground-state estimation in three sizes (groundstate). Table~\ref{tab:mqtcircuits} lists all families. Depth and gate counts are taken from the circuit files after removing measurements and barriers. Depth is the number of gate layers when each gate is placed directly after the last gate on any of its qubits.

\begin{table}[htbp]
\centering
\footnotesize
\setlength{\tabcolsep}{4pt}
\caption{MQT Bench circuit families. Qubits, Depth, and Gates give ranges over the circuits of each family, with depth and gate counts taken from the circuit files without measurements and barriers. Each circuit yields one amplitude network, two expectation networks ($\langle Z_c\rangle$ and $\langle Z_cZ_{c+1}\rangle$), and one state network, so a family with $k$ circuits has at most $k$, $2k$, and $k$ networks in the three task columns. Each task entry is the number of planned networks, which have 3 to 4999 tensors after simplification. A dash means that no network of the family is planned for that task.}
\label{tab:mqtcircuits}
\begin{tabular}{@{}lrrrrrrr@{}}
\toprule
Family & Qubits & Circuits & Depth & Gates & Ampl. & Expect. & State \\
\midrule
\multicolumn{8}{@{}l}{\emph{Families at every qubit count; $n=10,20,\ldots,130$}} \\
\quad ae & 10--130 & 13 & 53--773 & 91--9031 & -- & 26 & 13 \\
\quad dj & 10--130 & 13 & 11--131 & 28--388 & -- & -- & 13 \\
\quad ghz & 10--130 & 13 & 10--130 & 10--130 & -- & -- & 13 \\
\quad graphstate & 10--130 & 13 & 6--9 & 20--260 & -- & -- & 13 \\
\quad qft & 10--130 & 13 & 20--260 & 60--8580 & -- & -- & 13 \\
\quad qftentangled & 10--130 & 13 & 22--262 & 70--8710 & -- & -- & 13 \\
\quad qnn & 10--130 & 13 & 94--1294 & 319--50959 & 7 & 10 & 7 \\
\quad qpeexact & 10--130 & 13 & 24--388 & 66--8708 & -- & 12 & 13 \\
\quad qpeinexact & 10--130 & 13 & 28--388 & 68--8708 & -- & 26 & 13 \\
\quad random & 10--130 & 13 & 55--1803 & 191--39761 & 6 & 8 & 6 \\
\quad realamprandom & 10--130 & 13 & 41--521 & 175--25675 & 5 & 8 & 5 \\
\quad su2random & 10--130 & 13 & 41--521 & 175--25675 & 5 & 8 & 5 \\
\quad twolocalrandom & 10--130 & 13 & 41--521 & 175--25675 & 5 & 8 & 5 \\
\quad wstate & 10--130 & 13 & 20--260 & 37--517 & -- & -- & 13 \\
\addlinespace
\multicolumn{8}{@{}l}{\emph{Families at fixed sizes; all files}} \\
\quad groundstate & 4--14 & 3 & 12--42 & 24--224 & 3 & 3 & 3 \\
\quad grover-noancilla & 2--11 & 10 & 1--76676 & 2--77161 & 2 & 10 & 6 \\
\quad grover-v-chain & 2--17 & 10 & 1--51527 & 2--52011 & 4 & 10 & 6 \\
\quad portfolioqaoa & 3--17 & 15 & 13--69 & 21--476 & 14 & 28 & 14 \\
\quad portfoliovqe & 3--18 & 16 & 13--73 & 21--531 & 16 & 32 & 16 \\
\quad pricingcall & 5--25 & 11 & 35--7428 & 42--7500 & -- & 18 & 11 \\
\quad pricingput & 5--25 & 11 & 35--7434 & 42--7526 & -- & 18 & 11 \\
\quad qaoa & 3--16 & 14 & 8--10 & 15--80 & 13 & 26 & 13 \\
\quad qwalk-noancilla & 3--13 & 11 & 19--98131 & 25--98325 & -- & 8 & 6 \\
\quad qwalk-v-chain & 3--31 & 15 & 19--1363 & 25--1453 & -- & 28 & 15 \\
\quad routing & 2--12 & 3 & 7--19 & 11--81 & 1 & 4 & 2 \\
\quad shor & 18 & 2 & 6213 & 9875 & 1 & -- & -- \\
\quad tsp & 4--16 & 3 & 17--29 & 39--171 & 3 & 6 & 3 \\
\quad vqe & 3--16 & 14 & 7--20 & 13--78 & -- & 26 & 13 \\
\midrule
Total & & 320 & & & 85 & 323 & 264 \\
\bottomrule
\end{tabular}
\end{table}

\subsection{Network construction}
For each circuit, we remove final measurements, barriers, and classical registers. No selected circuit contains a reset or a gate after a measurement. quimb~1.15 reads the circuit in \texttt{complex128}. Because quimb does not support the \texttt{rzz}, \texttt{rxx}, \texttt{ch}, and \texttt{rccx} gates used by some files, we replace the first three by exact decompositions and \texttt{rccx} by its definition in \texttt{qelib1.inc}, and we checked each replacement against the gate matrix.  For the state task with $n>28$, the open qubits are $q_{\mathrm{round}(i(n-1)/19)}$ for $i=0,\ldots,19$. We then apply quimb's \texttt{full\_simplify} with the sequence ADCRS (antidiagonal gauging, diagonal reduction, column reduction, rank simplification, and split simplification), tolerance $10^{-12}$, and norm equalization.  Twenty-four networks, mostly amplitudes of Grover, quantum-walk, and Shor circuits, evaluate exactly to zero. Norm equalization is undefined for them, so we build them without it, and simplification then reduces each to a single tensor. Each network is built in a separate process with a 3600-s limit. The exported networks keep indices shared by more than two tensors. On small circuits, we verified that contracting the exported networks reproduces the amplitudes, expectation values, and states computed from quimb's dense state vector.

\subsection{Run protocol}
All runs used the server of Section~\ref{sec:setup}, which has eight CPU sockets with 16 physical cores each, and the software versions of Table~\ref{tab:versions}. A scheduler ran up to 16 planner processes, each pinned to eight physical cores of one socket, with at most eight TNCO processes at a time, and started the networks in order of increasing tensor count. Planning time is measured from entry into the planner API until it returns, and the common evaluator recomputes all metrics from the returned order on the simplified network. Cotengra checks its time limit between trials, so 45 Hyper128 runs took 3501 to 3634~s to return their orders. We report their planning times as measured and include these runs in all comparisons.

\subsection{Complete results}
Tables~\ref{tab:mqtresults} and~\ref{tab:mqttasks} report all four cost measures and planning time by network size and by task.

\begin{table}[htbp]
\centering
\footnotesize
\setlength{\tabcolsep}{7pt}
\caption{MQT Bench results relative to \emph{Heavy}, by network size. Entries are geometric means, over networks, of each configuration's value divided by that of \emph{Heavy}, so values above one favor \emph{Heavy}. Writes, Largest, and Peak denote $W(P)$, $\Mmax(P)$, and $L_{\max}(P)$, and Time is planning time. The \emph{Heavy} rows equal one by definition. Parentheses give the number of networks and, for the size classes, their number of tensors after simplification.}
\label{tab:mqtresults}
\begin{tabular}{@{}lrrrrr@{}}
\toprule
Configuration & FLOPs & Writes & Largest & Peak & Time \\
\midrule
\multicolumn{6}{@{}l}{\emph{All networks (672 in total)}} \\
\quad \arctn{} \emph{Heavy} & 1.00 & 1.00 & 1.00 & 1.00 & 1.00 \\
\quad \arctn{} \emph{Light} & 1.12 & 1.10 & 1.19 & 1.17 & 0.39 \\
\quad Cotengra Hyper128 & 1.04 & 1.10 & 1.10 & 1.06 & 52.7 \\
\quad Cotengra AutoHQ & 1.07 & 1.11 & 1.02 & 0.98 & 1.53 \\
\quad Cotengra Auto & 34.3 & 35.0 & 39.7 & 34.5 & 0.45 \\
\quad OMECO TreeSA-8 (50 iter.) & 11.8 & 12.4 & 13.7 & 13.9 & 0.51 \\
\quad OMECO TreeSA-8 (200 iter.) & 5.08 & 5.17 & 5.59 & 5.72 & 1.91 \\
\quad TNCO & 207 & 88.0 & 154 & 128 & 0.33 \\
\quad TNCO (\texttt{fuse=4}) & 42.6 & 23.1 & 38.6 & 31.7 & 0.33 \\
\addlinespace
\multicolumn{6}{@{}l}{\emph{Small networks (530 networks, 3--999 tensors)}} \\
\quad \arctn{} \emph{Heavy} & 1.00 & 1.00 & 1.00 & 1.00 & 1.00 \\
\quad \arctn{} \emph{Light} & 1.07 & 1.04 & 1.10 & 1.10 & 0.36 \\
\quad Cotengra Hyper128 & 0.99 & 1.00 & 0.94 & 0.93 & 61.0 \\
\quad Cotengra AutoHQ & 1.01 & 1.00 & 0.95 & 0.93 & 1.26 \\
\quad Cotengra Auto & 2.51 & 2.48 & 2.67 & 2.42 & 0.64 \\
\quad OMECO TreeSA-8 (50 iter.) & 1.08 & 1.02 & 1.11 & 1.09 & 0.59 \\
\quad OMECO TreeSA-8 (200 iter.) & 1.03 & 0.96 & 1.02 & 1.02 & 2.25 \\
\quad TNCO & 3.97 & 3.35 & 3.92 & 3.61 & 0.58 \\
\quad TNCO (\texttt{fuse=4}) & 2.69 & 2.39 & 2.96 & 2.57 & 0.58 \\
\addlinespace
\multicolumn{6}{@{}l}{\emph{Large networks (142 networks, 1000--4999 tensors)}} \\
\quad \arctn{} \emph{Heavy} & 1.00 & 1.00 & 1.00 & 1.00 & 1.00 \\
\quad \arctn{} \emph{Light} & 1.29 & 1.36 & 1.59 & 1.47 & 0.51 \\
\quad Cotengra Hyper128 & 1.23 & 1.53 & 1.94 & 1.69 & 30.7 \\
\quad Cotengra AutoHQ & 1.34 & 1.61 & 1.37 & 1.18 & 3.12 \\
\quad Cotengra Auto & $6.0\times10^{5}$ & $6.8\times10^{5}$ & $9.4\times10^{5}$ & $7.0\times10^{5}$ & 0.12 \\
\quad OMECO TreeSA-8 (50 iter.) & $8.8\times10^{4}$ & $1.3\times10^{5}$ & $1.6\times10^{5}$ & $1.8\times10^{5}$ & 0.29 \\
\quad OMECO TreeSA-8 (200 iter.) & $2.0\times10^{3}$ & $2.8\times10^{3}$ & $3.2\times10^{3}$ & $3.6\times10^{3}$ & 1.05 \\
\quad TNCO & $5.3\times10^{8}$ & $1.7\times10^{7}$ & $1.4\times10^{8}$ & $7.8\times10^{7}$ & 0.04 \\
\quad TNCO (\texttt{fuse=4}) & $1.3\times10^{6}$ & $1.1\times10^{5}$ & $5.7\times10^{5}$ & $3.8\times10^{5}$ & 0.04 \\
\bottomrule
\end{tabular}
\end{table}

\begin{table}[htbp]
\centering
\footnotesize
\setlength{\tabcolsep}{7pt}
\caption{MQT Bench results relative to \emph{Heavy}, by task. Columns are as in Table~\ref{tab:mqtresults}. Expectation combines $\langle Z_c\rangle$ and $\langle Z_cZ_{c+1}\rangle$.}
\label{tab:mqttasks}
\begin{tabular}{@{}lrrrrr@{}}
\toprule
Configuration & FLOPs & Writes & Largest & Peak & Time \\
\midrule
\multicolumn{6}{@{}l}{\emph{Amplitude (85 networks)}} \\
\quad \arctn{} \emph{Heavy} & 1.00 & 1.00 & 1.00 & 1.00 & 1.00 \\
\quad \arctn{} \emph{Light} & 1.17 & 1.12 & 1.23 & 1.22 & 0.36 \\
\quad Cotengra Hyper128 & 1.72 & 1.78 & 2.10 & 1.98 & 70.9 \\
\quad Cotengra AutoHQ & 1.82 & 1.93 & 1.87 & 1.77 & 3.58 \\
\quad Cotengra Auto & 23.4 & 23.8 & 30.2 & 26.5 & 0.45 \\
\quad OMECO TreeSA-8 (50 iter.) & 3.03 & 2.98 & 3.37 & 3.66 & 0.56 \\
\quad OMECO TreeSA-8 (200 iter.) & 2.10 & 2.05 & 2.26 & 2.48 & 2.14 \\
\quad TNCO & 10.7 & 3.49 & 7.02 & 6.08 & 0.29 \\
\quad TNCO (\texttt{fuse=4}) & 6.74 & 2.54 & 5.45 & 4.38 & 0.30 \\
\addlinespace
\multicolumn{6}{@{}l}{\emph{Expectation (323 networks)}} \\
\quad \arctn{} \emph{Heavy} & 1.00 & 1.00 & 1.00 & 1.00 & 1.00 \\
\quad \arctn{} \emph{Light} & 1.13 & 1.13 & 1.26 & 1.21 & 0.41 \\
\quad Cotengra Hyper128 & 0.80 & 0.86 & 0.80 & 0.76 & 61.1 \\
\quad Cotengra AutoHQ & 0.84 & 0.87 & 0.73 & 0.69 & 3.61 \\
\quad Cotengra Auto & 262 & 269 & 314 & 250 & 0.33 \\
\quad OMECO TreeSA-8 (50 iter.) & 96.5 & 107 & 123 & 121 & 0.44 \\
\quad OMECO TreeSA-8 (200 iter.) & 20.7 & 21.6 & 23.7 & 23.8 & 1.68 \\
\quad TNCO & $3.3\times10^{3}$ & $1.3\times10^{3}$ & $2.9\times10^{3}$ & $2.1\times10^{3}$ & 0.20 \\
\quad TNCO (\texttt{fuse=4}) & 414 & 236 & 479 & 343 & 0.19 \\
\addlinespace
\multicolumn{6}{@{}l}{\emph{State (264 networks)}} \\
\quad \arctn{} \emph{Heavy} & 1.00 & 1.00 & 1.00 & 1.00 & 1.00 \\
\quad \arctn{} \emph{Light} & 1.09 & 1.05 & 1.09 & 1.10 & 0.37 \\
\quad Cotengra Hyper128 & 1.22 & 1.26 & 1.32 & 1.29 & 40.0 \\
\quad Cotengra AutoHQ & 1.22 & 1.25 & 1.28 & 1.23 & 0.40 \\
\quad Cotengra Auto & 3.23 & 3.26 & 3.44 & 3.33 & 0.63 \\
\quad OMECO TreeSA-8 (50 iter.) & 1.39 & 1.39 & 1.47 & 1.51 & 0.58 \\
\quad OMECO TreeSA-8 (200 iter.) & 1.21 & 1.21 & 1.28 & 1.31 & 2.16 \\
\quad TNCO & 18.3 & 9.24 & 11.6 & 11.2 & 0.63 \\
\quad TNCO (\texttt{fuse=4}) & 4.76 & 2.73 & 3.34 & 3.24 & 0.63 \\
\bottomrule
\end{tabular}
\end{table}

\end{document}